\documentclass[a4paper, onecolumn, 12pt]{IEEEtran}
\usepackage[cmex10]{amsmath}
\usepackage{amssymb}
\usepackage{amsthm}
\usepackage{array}
\usepackage{graphicx}
\usepackage{amssymb,amsmath}
\usepackage{dsfont}
\usepackage{array}
\usepackage{cite}
\usepackage{multirow}
\usepackage{algorithm}
\usepackage{algpseudocode}
\usepackage{algorithmicx}

\usepackage{float}
\newfloat{floatalgorithm}{t}{lop}
\floatname{floatalgorithm}{Algorithm}

\renewcommand{\qed}{\hfill$\square$}

\theoremstyle{example}
\newtheorem{definition}{Definition}
\usepackage{tikz}

\newtheoremstyle{example}{\topsep}{\topsep}{}{}{\itshape}{:}{.5em}{\thmname{#1}\thmnumber{ #2}\thmnote{ (#3&)}}
\newtheoremstyle{examplecontd}{\topsep}{\topsep}{}{}{\itshape}{:}{.5em}{\thmname{#1}\thmnumber{ #2}\thmnote{ #3&}\enspace(Cont'd)}
\theoremstyle{example}

\theoremstyle{example}
\newtheorem{theorem}{Theorem}
\newtheorem{lemma}{Lemma}
\newtheorem{corollary}{Corollary}

\def\remark{
  \let\go\relax
  \ifvmode\vskip-\lastskip\fi
  \noindent{\it Remark\/.}%
  \enskip\relax\ignorespaces\go}
\newcommand{\bs}[1]{\ensuremath{\boldsymbol{#1}}}

\newcommand{\bldc}{{\mbox{\boldmath $c$}}}
\newcommand{\bldB}{{\mbox{\boldmath $B$}}}
\newcommand{\bldH}{{\mbox{\boldmath $H$}}}
\newcommand{\bldP}{{\mbox{\boldmath $P$}}}
\newcommand{\bldW}{{\mbox{\boldmath $W$}}}

\newcommand{\GF}{{\mbox{GF}}}

\begin{document}

%\title{On optimizing LDPC code performance}
\title{BP-LED decoding algorithm for LDPC codes  over AWGN channels }
%\author{
%\IEEEauthorblockN{Irina E. Bocharova$^1$,  %Florian Hug$^2$,
%Rolf Johannesson$^2$, and Boris D. Kudryashov$^1$}
%\vspace{1mm}
%\IEEEauthorblockA{
%	\small
%	\begin{tabular}{c@{\hspace{1.5cm}}c}
%		$^1$ Dept. of Information Systems                 &  $^2$ Dept. of Electrical and Information Technology,\\
%		St. Petersburg Univ. of Information Technologies, &  Lund University\\
%		Mechanics and Optics                              &  P. O. Box 118, SE-22100 Lund, Sweden\\
%		St. Petersburg 197101, Russia                     &  Email: rolf@eit.lth.se\\
%		Email: \{irina, boris\}@eit.lth.se
%	\end{tabular}
%	\vspace{-4mm}
%}
%}
\author{
\IEEEauthorblockN{{\bf Irina E. Bocharova$^{1,2}$, 
 Boris D. Kudryashov$^1$, Vitaly Skachek$^2$, \\ and
Yauhen Yakimenka$^2$}}
\vspace{1mm}
\IEEEauthorblockA{
	\small
	\begin{tabular}{c@{\hspace{1.5cm}}c}
		$^1$ Department of Information Systems                 &  $^2$ Institute of Computer Science \\
		St. Petersburg University of Information Technologies, &  University of Tartu\\
		Mechanics and Optics                                &  Tartu 50409, Estonia\\
		St. Petersburg 197101, Russia                       &  Email:  \{ vitaly, yauhen \} @ut.ee\\
 		Email: irinaboc@ut.ee,  kudryashov\_boris@bk.ru								&  
	\end{tabular}
	\vspace{-4mm}
}
}
\maketitle
\begin{abstract}
A new method for low-complexity near-maximum-likelihood
(ML) decoding of low-density parity-check (LDPC)
codes over the additive white Gaussian noise channel is presented. The proposed method termed belief-propagation--list erasure decoding (BP-LED)  is based 
on erasing carefully chosen unreliable bits performed  in case of BP decoding failure. A strategy of introducing erasures into the received vector and a new erasure decoding algorithm are proposed. The  new erasure decoding algorithm, called list erasure decoding, combines  ML decoding over the BEC with list decoding applied if the ML decoder fails to find a unique solution.  The asymptotic exponent of the average list size for random regular  LDPC codes from the Gallager ensemble is analyzed.  Furthermore, a few examples of regular
and irregular quasi-cyclic LDPC codes of short and moderate lengths  are studied by simulations
and their performance is compared with the upper bound  on the LDPC ensemble-average performance and the upper bound on the average performance of random linear codes under ML decoding. A comparison with the BP decoding performance of the WiMAX standard codes and performance of the near-ML BEAST decoding are presented. The new algorithm is applied to decoding a short nonbinary  LDPC code over the extension of the binary Galois field. The obtained simulation results are compared to the upper bound on the ensemble-average performance of the binary image of regular nonbinary LDPC codes.\footnote{Results of this work were partly published in\cite{bocharova2016low}. This work is supported by the Norwegian-Estonian Research Cooperation Programme through the grant EMP133.} 
\end{abstract}

%%%%%%%%%%%%%%%%%%%%%%%%%%%%%%%%%%%%%%
\section{Introduction}\label{sec:1}
%%%%%%%%%%%%%%%%%%%%%%%%%%%%%%%%%%%%%%
Since their rediscovery in 1995, low-density parity-check (LDPC) codes in conjunction with iterative decoding continue to attract attention of both researches and companies developing communication standards. The main reason for this popularity of LDPC codes is their near-Shannon limit performance. Although there exist asymptotic ensembles of LDPC codes approaching  capacity under belief propagation (BP) decoding,  performance of finite length LDPC codes under  BP decoding is inferior to their performance under maximum-likelihood (ML) decoding. Moreover, the performance gap between  ML and BP decoding increases when signal-to-noise ratio (SNR) grows. Typically, LDPC codes under   BP decoding suffer from the so-called \emph{error floor} phenomen, which is caused by both sub-optimality of the decoding algorithm and by structural properties of the codes.    

There exists a variety of techniques for lowering the error floors, and they are usually based on identifying and removing specific structural configurations of the code Tanner graph called trapping sets~\cite{Richardson2003}. This can be done by modifying both the code parity-check matrix (without changing the code) and the iterative decoding algorithm.  For example, in~\cite{cavus}, a list of trapping sets is computed and stored in a look-up table. If the BP decoder fails, then a post-processing based on the list of known trapping sets is performed. Techniques combining the trapping set detection with code shortening or bit-pinning are presented in~\cite{han2009low} and~\cite{zhangryan2009}. 
A technique based on eliminating small trapping sets by adding extra checks to the code parity-check matrix is studied in \cite{mu2011}. A method for eliminating small trapping sets when constructing an LDPC code is considered in \cite{asvadi2011lowering}. The proposed method is applicable to both regular and irregular LDPC codes.

Another approach for improving the performance of  BP decoding stems from the information set 
decoding. The most efficient methods for near-optimal decoding of linear codes are 
based on multiple attempts for finding an error-free information set and 
subsequent reconstruction of a codeword by re-encoding~\cite{prange1962use, valembois2004box, fossorier1995soft}. In~\cite{pishro2007results} and~\cite{varnica}, a post-processing in the form of bit-guessing is applied to the output of the BP decoder in case of its failure. In particular, in \cite{pishro2007results}, the bits that participate in the largest number of unsatisfied checks are guessed, and  BP decoding is repeated for each guessing attempt. 
An improved algorithm for selecting the bits to be guessed is suggested in \cite{varnica}. There, the post-processing step is performed in stages. That technique leads to better than in \cite{pishro2007results}  performance of near-ML decoding at the cost of higher computational complexity. A method for decoding of nonbinary LDPC codes over extensions of the binary Galois field based on the ordered statistics approach \cite{fossorier1995soft} is studied in~\cite{baldi2014hybrid}. 

Reconstruction of a codeword from a susbet of its symbols is equivalent to decoding 
over a binary erasure channel (BEC). It was shown in \cite{zyablov1974decoding}
 that  ML decoding of an $[n,k]$ LDPC code (where $n$ is the code length  and $k$ is the number of the information symbols) with $\nu = \Theta(n)$ erasures is equivalent to 
solving a system of linear equations of order $\nu$, that is, it can be performed via the Gaussian elimination with time complexity at most $O(\nu^3)$. By taking into account the sparsity of the parity-check matrix of the codes, the complexity can be lowered to approximately $O(\nu^2)$ (see overview and analysis in \cite{burshtein2004efficient} and the references therein).  
Practically feasible algorithms with thorough complexity analysis can be found in   
\cite{paolini2012maximum}. % liva
 
 Low-complexity suboptimal decoding techniques for LDPC codes over a BEC are often based on the following two approaches. In the first approach, redundant rows and columns are added to the original code parity-check matrix (see, for example, %\cite{kasai2004},
 \cite{vasic2005},\cite{kobayashi2006}). In the second approach, a post-processing is 
used in case BP decoding fails \cite{pishro2004decoding,hosoya2004,olmos2010tree}. 

The idea to reduce the problem of the decoding of an LDPC code over the AWGN channel to the 
decoding of erasures has first appeared in \cite{fang2010bp}. In that work,  BP decoding is followed by introducing artificial erasures and  their subsequent  decoding over the BEC, thus yielding a near-ML decoding algorithm. The performance of the decoding algorithm in \cite{fang2010bp} strongly depends on the efficiency of the procedure that converts the AWGN channel into the BEC  or, in other words, the procedure for selecting the bits, which are to be erased. The channel transformation can be considered successful only if non-erased bits of the input vector are error-free.  

In this paper, we propose a new decoding algorithm, which uses the ideas 
similar to \cite{fang2010bp} and \cite{pishro2007results}, but differs significantly
both in a strategy for introducing erasures and in an  erasure decoding algorithm. A new list erasure decoding (LED) algorithm, which is applied to the result of the BP decoding in case of its failure can, in principle, be combined with any type of soft decoding on the AWGN channel.  A distinguishing feature of the LED algorithm is an additional search step over a fixed size list of unresolved bit positions which is applied if the ML decoder over the BEC fails to find a unique solution.   
 %The second algorithm after selecting bits to be erased  additionally uses error-erasure separating  \cite{abdelghafar2013}  based on shortening of the extended code parity-check matrix.  
The proposed algorithm which we call BP-LED decoding is tested on the regular quasi-cyclic (QC) LDPC code with optimized girth of the code Tanner graph~\cite{bocharova2011double} and on the irregular QC LDPC codes optimized by using the technique in~\cite{Boch2016}. Both binary LDPC codes and binary images of nonbinary LDPC codes are studied.  Simulation results are presented. 
The comparison with the theoretical bounds on the ML decoding performance is performed.

The rest of the paper is organized as follows. Some necessary definitions are given in Section \ref{sec_Prelim} and known bounds on the error probability of ML decoding are revisited in Section \ref{sec_Bounds}. List-decoding algorithm over the BEC is described in Section \ref{BEC} and its analysis for the Gallager ensemble of regular LDPC codes is presented in Section \ref{sec_Analysis}. Techniques for selecting bit positions to be erased are discussed in Section \ref{Selection}. 
The near ML decoding procedure using LED is described in Section \ref{ledbased}.
The paper is concluded by the discussion of the simulation results and their comparison with the bounds on the error probability of  ML decoding in Section \ref{sec_Discussion}.
%\footnote{\vit{What about Section VII?}}

\section{Preliminaries}
\label{sec_Prelim}
%In this paper, we propose a new algorithm for near-ML decoding of both regular and irregular LDPC codes. We also 
%present analysis of this algorithm for the ensemble of random regular LDPC codes. The simulation results are compared to the %theoretical bounds on the error probability of the ML decoding of the random linear codes and of the random regular LDPC codes.

Consider  the Gallager ensemble of $(J,K)$-regular LDPC codes of length $n$ and dimension $k$ \cite{gallager}. In this ensemble, an $r \times n$ random parity-check matrix $\bldH$ that consists of $J$ strips $\bldH_{i}$ of width $M=r/J$ rows each,  $ i =1,2,\dots, J$, where $r=n-k$. All strips are random column permutations of the strip  where the $j$th row contains $K$ ones in positions $(j-1)K+1, (j-1)K+2, \ldots, jK$, for $j = 1, 2, \ldots, n/K$.  %Since we are interested in performance of near-ML decoding the average minimum distance over the Gallager ensemble is optimized over $J$ \cite{ISIT17average}. Then among randomly generated $(J,K)$-LDPC codes with optimized $J$ value we search for the codes with the largest minimum distance. 

%Interpreting  $\bldH$ as  a biadjacency matrice \cite{biadj}
%yields its corresponding Tanner graph.  The girth $g$  is  the length of the shortest cycle 
%of the code Tanner graph.

%The QC LDPC codes are very often used in practice. The decoding algorithm, which is proposed in the sequel, is suitable for use %with any LDPC code, yet we apply it specifically to QC LDPC codes.
Rate $R=b/c$  QC LDPC codes are determined  by a $(c-b)\times c$ polynomial parity-check matrix of their parent convolutional code \cite{johannesson2015fundamentals}
\begin{IEEEeqnarray}{C}
\bldH(D)=\left(\begin{array}{cccc}
h_{11}(D)&h_{12}(D)&\dots&h_{1c}(D)\\
h_{21}(D)& h_{22}(D)&\dots&h_{2c}(D)\\
\vdots&\vdots&\ddots&\vdots\\
h_{(c-b)1}(D)& h_{(c-b)2}(D)&\dots&h_{(c-b)c}(D)
\end{array}
\right)\IEEEeqnarraynumspace
\label{polynom_matr}
\end{IEEEeqnarray}
where $h_{ij}(D)$ is  either zero or a monomial entry, that is, $h_{ij}(D)\in \{0,D^{w_{ij}}\}$
with $w_{ij}$ being a nonnegative integer, $w_{ij}\le \mu$, and $\mu=\max_{i,j} \{ w_{ij} \}$ is the syndrome memory.
%Its 
The polynomial matrix (\ref{polynom_matr})
can be represented via  
the $(c-b)\times c$  {\em degree matrix}  
\begin{equation} \label{defW}
\bldW=\begin{pmatrix}
w_{11}&w_{12}&\dots&w_{1c}\\
w_{21}& w_{22}&\dots&w_{2c}\\
\vdots&\vdots&\ddots&\vdots\\
w_{(c-b)1}& w_{(c-b)2}&\dots&w_{(c-b)c}
\end{pmatrix}
\end{equation}
with entries $w_{ij}$ at the positions of the monomials $h_{ij}(D)=D^{w_{ij}}$. 
We write $w_{ij}=-1$ for the positions where $h_{ij}(D)=0$.
By tailbiting the parent convolutional code to length $M > \mu$, we obtain
the binary parity-check matrix  (see \cite[Chapter 2]{johannesson2015fundamentals})
\begin{IEEEeqnarray}{C}
%H_{\rm TB}^{\rm T}=\left(\begin{array}{cccccc}
\bldH^{\rm T}=\begin{pmatrix}
\bldH_{0}^{\rm T}&\bldH_{1}^{\rm T}&\dots&\bldH_{\mu-1}^{\rm T}&\bldH_{\mu}^{\rm T}&{\bs 0}&\dots&\bs 0\\
{\bs 0}& \bldH_{0}^{\rm T}&\bldH_{1}^{\rm T}&\dots&\bldH_{\mu-1}^{\rm T}&\bldH_{\mu}^{\rm T}&\dots&\bs 0\\
\vdots &                 &\ddots               &\vdots&\vdots                     &\vdots                   &\ddots&\\
\bldH_{\mu}^{\rm T}& {\bs 0}&\dots&\bs 0&\bldH_{0}^{\rm T}&\bldH_{1}^{\rm T}&\dots&\bldH_{\mu-1}^{\rm T}\\
\vdots&\ddots&\vdots&\vdots&\vdots&\vdots&\vdots&\vdots\\
\bldH_{1}^{\rm T}&\dots&\bldH_{\mu}^{\rm T}&{\bs 0}&\dots&\bs 0&\dots&\bldH_{0}^{\rm
T}
\end{pmatrix}
%\right)\IEEEeqnarraynumspace
\label{tb}
\end{IEEEeqnarray}
of the $[Mc,Mb]$ QC LDPC block code of length $Mc$ and dimension $Mb$, where 
$\bldH_{i}$, $i=0,1,\dots,\mu$, are binary $(c-b)\times c$ matrices in the series expansion 
\begin{equation*}
	\bldH(D)=\bldH_{0}+\bldH_{1}D+\dots+\bldH_{\mu}D^{\mu}
\end{equation*}
and $\bs 0$ is the all-zero matrix of size $(c-b)\times c$. Further by $\bs 0$ we denote the all-zero matrix of an appropriate size. 
If each column of $\bldH(D)$ contains $J$  nonzero elements, and each row contains
$K$ nonzero elements the QC LDPC block code is $(J,K)$-regular. It is irregular otherwise.

Another form of the equivalent $[Mc,Mb]$ binary QC LDPC block code can be obtained by replacing
the nonzero monomial elements of $\bldH(D)$ in (\ref{polynom_matr})  by the  powers of the circulant $M\times M$  permutation matrix $\bldP$, whose rows are cyclic shifts by one position to the right of the rows of the identity matrix.

The polynomial parity-check matrix $\bldH(D)$ (\ref{polynom_matr}) can be interpreted as a $(c-b) \times c$ binary base matrix $\bldB$ labeled by monomials, where the entry in $\bldB$ is one if and only if the corresponding entry of $\bldH(D)$ is nonzero, i.e.
\[
\bldB=\bldH(D)|_{D=1}
\]

By viewing $\bldH$ as a biadjacency matrix \cite{graph}, 
we obtain a corresponding bipartite Tanner graph.  The girth $g$ is the length of the shortest cycle 
in the Tanner graph.

\section{Tightened Bounds on the ML decoding error probability }
\label{sec_Bounds}
In what follows, we compare  the performance of the new decoding algorithm with the performance of   ML decoding over an AWGN channel. While keeping that in mind,  in this section we revisit known bounds  on the error probability of  ML decoding over the  AWGN channel.  

By using technique in \cite{bocharova2017} we compute the exact spectrum coefficients for the Gallager ensembles of binary regular LDPC codes and binary images of nonbinary regular LDPC codes. By substituting the computed coefficients into the existing bounds on the error probability of  ML decoding, we obtain new bounds, which are  tighter than the previously known counterparts.  

\subsubsection{Lower bound} 

Since 1959, the Shannon bound \cite{Shannon1959} is still the best known lower bound on the ML  decoding error probability for codes used over the  AWGN channel in a wide range of rates and lengths \cite{sason2006performance}.
Computational aspects of this bound are studied in \cite{valembois2004sphere} (see also \cite{sason2006performance} for
overview of the results in this area). 
In the sequel, we use approximation in~\cite{BP2006} of the Shannon bound \cite{Shannon1959}
which gives values indistinguishable 
from the values of the bound in~\cite{Shannon1959} for the frame error rate (FER) performance below 0.1 over the AWGN channel.

Let $n$, $R$, and $\sigma$  denote the  code length, code rate and  standard noise deviation for an AWGN channel, respectively. We use notations and formulas in  \cite{Shannon1959} for the cone half-angle $\theta\in [0,\pi]$, which
corresponds to the solid angle of an $n$-dimensional circular cone, and for the solid angle of the whole space
\[
\Omega_{n}(\theta)=\frac{2\pi^\frac{n-1}{2}}{\Gamma(\frac{n-1}{2})}\int_{0}^{\theta}(\sin \phi)^{n-2}d\phi, 
\;\;\;
\Omega_{n}(\pi)=\frac{2\pi^{n/2}}{\Gamma(n/2)} \;,
\]
respectively.
For a given code of length $n$ and cardinality $2^{nR}$, the parameter $\theta_0$ is selected as a solution of the equation
\[
\frac{\Omega_{n}(\theta_{0})}{\Omega_{n}(\pi)}=2^{-nR}.
\]
Then, for the FER  $P_{\rm sh}(n,R,\sigma)$, we use approximation in~\cite{BP2006}
%\begin{multline}
\begin{equation}
P_{\rm sh}(n,R,\sigma) \approx  \frac{1}{\sqrt{n\pi}} \cdot
\frac{1}{\sqrt{1+G^2}\sin\theta_{0}} \\
\cdot
\frac{\left[G\sin\theta_{0}\exp\left(-\frac{1}{2\sigma^2}+\frac{G}{2\sigma}
\cos \theta_{0}\right)\right]^{n}}
{ {\frac{G}{\sigma}\sin^{2}\theta_{0}-\cos\theta_{0}}} \; , 
\label{Shannon}
\end{equation}
%\end{multline}

where $
G=\frac{1}{2\sigma}\left(\cos\theta_0+\sqrt{\cos^{2}\theta_0+4\sigma^2}\ \right)
$.

\subsubsection{Upper bound}
%\irina{
The tangential sphere bound (TSB) (also known as the Poltyrev upper bound \cite{poltyrev1994bounds}) is based on Gallager's bounding technique. Given a transmitted vector, the decoding FER is represented in the form
\begin{eqnarray}
{P}_{e} & = & {\rm Pr}(e,{\bs r}\in \mathcal{ R}) + {\rm Pr}(e,{\bs  r} \notin \mathcal{ R})  \nonumber\\
& = & {\rm Pr}(e, {\bs r} \in\mathcal { R}) + {\rm Pr}(e|{\bs r}\notin {\mathcal  R}){\rm Pr}({\bs r}\notin {\mathcal  R}) \nonumber\\
& \le &{\rm  Pr}(e,{\bs r}\in {\mathcal R}) +{\rm Pr}(\bs r \notin \mathcal R) \label{firstbound} , 
\end{eqnarray}
where $e$ is a decoding error event, $\bs  r$ is the received vector and $\mathcal R$ denotes the region,  whose choice significantly influences
the tightness of the bound  (\ref{firstbound}).
%}
%\irina{
The tightest bound in~\cite{poltyrev1994bounds} uses a conical region $\mathcal  R$. It combines the
ideas of spherical approach  \cite{hughes1991error}, which considers a spherical regions $\mathcal R$,  and the tangential bound  \cite{berlekamp1980technology}, which decomposes the noise vector into the radial and tangential components.
%}
We present here the Poltyrev  bound \cite{poltyrev1994bounds} for completeness: 
\begin{eqnarray}
P_e &\le&  \int_{-\infty}^{\sqrt n} f\left(\frac{x}{\sigma}\right)
\left\{
\sum_{w\le w_0} S_w \Theta_w(x)
+
 1-\chi_{n-1}^2\left(\frac{r^2_x}{\sigma^2}\right)
\right\}dx + Q\left( \frac{\sqrt n}{\sigma}\right).
\label{TSB}
\end{eqnarray}
Here $f(x)=\frac{1}{\sqrt{2\pi}}\exp{\{-x^2/2\}}$ is the  Gaussian probability density function,  
$Q(x)  =  \int_{x}^{\infty} f(x) dx$, 
\begin{eqnarray*}
\Theta_w(x) & = & 
 \int_{\beta_w(x)}^{r_x}
f\left(\frac{y}{\sigma}\right)\chi_{n-2}^2
\left(\frac{r^2_x-y^2}{\sigma^2}\right)
dy \; , 
\\
w_0&=&\left \lfloor \frac{r_0^2n}{r_0^2+n}\right \rfloor , \qquad
r_x=r_0\left(1-\frac{x}{\sqrt n}\right) \; , \qquad 
%\mu_w(r)&=&\frac{1}{r}\sqrt {\frac{w}{1-w/n}}, 
\beta_w(x)=\left(1-\frac{x}{\sqrt n}\right)\sqrt {\frac{w}{1-w/n}} \; , 
\end{eqnarray*}
$S_{w}$ is the $w$-th  spectrum coefficient of the code weight spectrum, $n$ is the code length, and $\chi_{n}^2$ denotes the probability density function of chi-squared distribution  with $n$ degrees of freedom.

Parameter $r_0$ is a solution with respect to $r$ of the equation

\begin{equation}\label{TSB_eq}
\sum_{w \; : \; \mu_w(r)<1}S_w \int_0^{\arccos \mu_w(r)} \sin^{n-3}\phi \; d \phi =\sqrt{\pi} \cdot
\frac{\Gamma\left(\frac{n-2}{2}\right)}{\Gamma\left(\frac{n-1}{2}\right)},\; 
\end{equation}

\[
\mu_w(r)=\frac{1}{r} \cdot \sqrt{\frac{w}{1-w/n}}.
\]
We note  that in order to use the Poltyrev bound~(\ref{TSB}), one has  to know the weight spectrum of the code.  

In what follows,  we use the exact coefficients  of the average spectrum of the Gallager ensemble of LDPC codes computed by the recurrent procedure presented in \cite{bocharova2017}.  
%%%%%%%%%%%%%%%%%%%%%%%%%%%%%%%%%
Let \[\bldH=
\left(
\begin{array}{c}
\bldH_{1}\\
... \\
\bldH_{J}
\end{array}
\right)
\] be a parity-check matrix of an LDPC code randomly chosen  from the Gallager ensemble.
In our derivations we use the generating function of a sequence of code weight enumerators. In a general case, the generating function $f(s)$ for a sequence  of numbers $a_{0},a_{1},....$  is defined as follows:
\[f(s)=\sum_{n=0}^{\infty}a_{n}s^{n},\]
where $s$ is a formal variable. 

The  generating function of  
the number of binary sequences $\bs x$ of weight $w$ and length $n$   
satisfying the equality 
\begin{equation}
\bs x \bldH_{i}^{\rm T}=\bs 0, 
\label{Gal_ens}
\end{equation} where $i  \in \{1,2,\dots,J\}$,
is given by  
\begin{equation}
G(s)=\sum_{w=0}^{n}G_{n,w}s^{w}= \big( g(s) \big)^{M} \; , 
\label{genf} 
\end{equation}
where  
%$N_{n,w}$ is the number of binary sequences $\bs x$ of weight $w$ and length $n$   
%satisfying the equality $\bs x \bldH_{i}^{\rm T}=\bs 0$, $i  \in \{1,2,\dots,J\}$, 
%where $\bldH_{i}$ is the $i$th strip of $\bld\bldH$ consisting of $M$ rows,
$g(s)=\sum_{i=0}^{K}g_{i}s^{i}=\left((1+s)^{K}+(1-s)^{K}\right)/2$,
 $g_{i}=\binom{K}{i}$  if  $i$ is even, and $g_{i}= 0$ otherwise.

From (\ref{genf}), we obtain the recurrent relation

\begin{eqnarray}
G_{1,i}\!\!&\!\!=\!\!&\!g_{i}\mbox{, } \quad i=0,1,...,K \; , \label{composite1} \\
G_{j,w}\!\!&\!\!=\!\!&\!\!\sum_{i=0}^{K}g_i \cdot G_{j-1,w-i} \, , \quad
j=2,\ldots,M, \quad w=0,1,\ldots,jK\label{composite2}\,.
\end{eqnarray}

The probability that (\ref{Gal_ens}) is valid for a random $\bs x$  of length $n$ and weight $w$ is equal to
\begin{equation}p(w)=\frac{G_{n,w}}{\binom{n}{w}} \label{prob}\end{equation} for each of the matrices $\bldH_{i}$.
The average number of codewords of length $n$ and weight $w$ is  given by 
\begin{equation} {\rm E} \{A_{n,w} \}=\binom{n}{w}\big( p(w)\big)^{J}=\binom{n}{w}^{1-J}G_{n,w}^{J} \; ,\label{average} \end{equation}
where $\big( p(w) \big)^{J}$ is the probability that $\bs x$ satisfies (\ref{Gal_ens}) for all $i=1,2,\dots,J$ simultaneously,
and {\rm E}\{ $\cdot$ \} denotes the expected value of a random variable.

Similarly, average spectra for binary images of nonbinary LDPC codes over $GF(2^{m})$, $m\ge 2$  can also  be computed via the generating function 
\begin{eqnarray}
G(s)&=& F(\rho)\left|_{\rho=\phi(s)} \right. \; , \\ 
F(\rho)&=&\sum_{w=0}^{KM}F_{w}\rho^{w}=f(\rho)^{M} \; ,  
\label{genff} 
\end{eqnarray}
where $f(\rho)=\left((1+(q-1)\rho)^{K}+(q-1)(1-\rho)^{K}\right)/q $  (see \cite[Chapter 5]{gallager})
and 
\begin{eqnarray}
\psi(s)&=&\sum_{i=1}^{m}\psi_{i}s^{i}=\frac{(1+s)^{m}-1}{q-1} \; ,\\
\psi_{i}&=&\frac{1}{q-1}\binom{m}{i} \;  \label{phi}\quad 
\end{eqnarray}
%In order to simplify the computation of the  coefficients $f_i$  in the series expansion  $f(\rho)=\sum_{i=0}^{K} %f_i\rho^i$, we use the following recursion %$\alpha_{0}=1$
%%$\alpha_{0}&=&1 $
%\begin{eqnarray}
%\alpha_{0}&=&1,\;\alpha_{i}=(q-1)^{i-1}-\alpha_{i-1} \label{alpha1} \; , \\
%f_i&=&\binom{K}{i}\alpha_i \; . \label{alpha2}
%\end{eqnarray}
(see \cite{bocharova2017} for details).   

%The average spectrum coefficients are obtained  as
%\begin{equation} {\rm E} \{A_{n,w} \}=\binom{n}{w} p(w)^{J}
%=\binom{n}{w}^{1-J}G_{n,w}^{J} \; ,\label{nonbin} 
%\end{equation}
%where  
%$
%p(w)={\binom{n}{w}}^{-1}{G_{n,w}}, n=mKM \; , 
%$  and for $G_{n,w}$ we can either use~(\ref{Step41})--(\ref{Step42}) in the general 
%case, or a simpler recursion (\ref{composite1})--(\ref{composite2}) for the binary codes.
%The computational complexity  is determined by the main step in the recursion %(\ref{Step42}), 
%and it is proportional to $M \cdot Km =n$.

%%%%%%%%%%%%%%%%%%%%%%%%

\section{List Decoding over a BEC}
\label{BEC}
Let $\bldH=\left(\bs h_{1},\bs h_{2},\dots,\bs h_{n}\right)$ be an {$r \times n$} parity-check matrix of a binary linear $[n,k, d_{\min}]$ block code, {$r = n - k$},  where  ${\bs h}_i$ denotes the $i$-th column of $\bldH$. 
We use notation $\bldH_I$ for the 
submatrix of $\bldH$, whose columns are indexed by the set $I\subseteq\{1,2,...,n\}$. 

Consider  a BEC with erasure probability $\varepsilon > 0$.
The ML decoder  corrects any pattern of $\nu$ erasures if
$\nu\le  d_{\min}-1$.  If $d_{\min}\le \nu \le n-k $ then 
the ML decoder can correct some  erasure patterns. 
The number of such correctable patterns depends on the code structure. 

Let  $\bs y=(y_{1},y_{2}, \dots,y_{n})$ be a received vector, where $y_{i}\in\{0,1,\phi\}$, and the symbol $\phi$ represents erasures. 
We denote by  $\bs e = (e_{1},e_{2}, \dots,e_{n})$ a binary vector, such that 

\[e_{i}=\left\{ 
\begin{array}{cl}
1&\mbox{if } y_i = \phi\\
0&\mbox{if }y_i \in \{0,1\}
\end{array}
\right.
\]
for all $i =1, 2, \ldots, n$.  Let  $I(\bs e)$ be a set of nonzero coordinates of $\bs e$, $|I(\bs e)|=\nu$, 
and 
$\bs z=(z_1,z_2,...,z_{\nu})$  be a vector of unknowns located in positions indexed by the set $I(\bs e)$.  Let $ \tilde {\bs y}$ be the  vector $\bs y$ with unknowns $z_{i}$ in positions $I(\bs e)$.
 
Consider a system of linear equations
%\begin{equation} \label{eq:system}
$\tilde{ \bs y} \bldH^{\rm T}=\bs 0$
%\; , 
%\end{equation}
which can be reduced to
\begin{equation} \label{eq:system}
%\bs y_{I^{\rm c}(\bs e)} \bldH^{\rm T}_{I^{\rm c}(\bs e)} =
\bs z \bldH^{\rm T}_{I(\bs e)} =\bs s(\bs e),   \;  
\end{equation}
where $\bs s(\bs e)=\bs y_{I^{\rm c}(\bs e)} \bldH^{\rm T}_{I^{\rm c}(\bs e)} $ is a syndrome vector computed using 
non-erased positions of $\bs y$ 
and \[ {I^{\rm c }(\bs e)=\{1,2,\dots,n\}\setminus  I(\bs e)}.\] 
%Assuming  that  
%there are no hard-decision errors in non-erased
%positions of $\bs y$,  we have 
%\begin{equation} \label{syndrome}
%\bs s(\bs e) = \bs y_{I(\bs e)} \bldH^{\rm T}_{I(\bs e)}=\bs z \bldH^{\rm T}_{I(\bs e)} \; . 
%\end{equation}
%
%\vit{
%Consider a linear system of equations\footnote{\vit{The system that one should solve is not (1) (because (1) has no unknowns), but rather the system (2). I changed this in the draft.}} 
%where the erased coordinates of ${\bs y}$ are treated as unknowns. }
The solution of (\ref{eq:system}) is unique if 
$%\[
\rho \triangleq {\rm rank } \bldH_{I(\bs e)} {= \nu} %\; , 
$, %\]
otherwise the full list $\mathcal{L}$ of candidate solutions contains $T=|\mathcal L|=2^L$ elements, where
$%\[
L = \nu - \rho. \;  
$%\]  

If the code rate $R$ approaches the BEC capacity $C=1-\varepsilon$, 
the typical number of erasures $\nu \approx n \varepsilon \to n (1-R) = n-k$. In that case, with high probability, 
the dimension $L$ of the linear space of solutions of (\ref{eq:system}) is positive.

\begin{figure}
\centering
\begin{tikzpicture}
[
scale=0.6,transform shape,
state/.style={rectangle,draw=black!50,fill=white!,thick,minimum size=6mm},
oper/.style={circle,draw=black!50,fill=white!,thick,minimum size=5mm},
arr/.style={<->,auto,>=stealth},
point/.style={circle,draw=black!50, fill=black!,thick,minimum size=0.5mm}
]

\node() at (-1.2,3) { \huge $ H_{I(\bs e)}=$};

\filldraw[draw=white, fill=gray!10!white] (6,0)-- (12,0)--(12,2)--(6,2);

\draw [very thick] (0.3,0)--(0,0)--(0,6.7)--(0.3,6.7);
\draw [very thick] (11.7,0)--(12,0)--(12,6.7)--(11.7,6.7);
\draw[dotted] (0,6)--(12,6);
\draw[very thick, dotted](6,2)--(12,2);

\draw[thick,dotted](1.5,4.5)--(3.5,2.5);
\draw[thick,dotted](5,1)--(5.3,0.7);

\node() at (6,6.4) { \huge $\bs 0$};

\node() at (0.5,5.5) {\huge $1$};
\node() at (1,5) {\huge $1$};
\node() at (4.5, 1.5) {\huge $1$};
\node() at (4, 2) {\huge $1$};

\node() at (5.7,0.3) {\huge $1$};
%\node() at (5,1) {\huge $1$};

\draw [arr,very thin](10.8,0) -- (10.8,6);
\node() at (11.2,3) {\huge $\rho$};

\draw [arr,very thin](0,4.) -- (6,4.);
\node() at (3.3,4.3) {\huge $\rho$};

\draw [arr,very thin](9.6,0) -- (9.6,2);
\node() at (10.2,1) {\huge $\rho_A$};

\draw [very thick, dotted](6,0)--(6,6);

\node() at (8,1) { \Huge $\bs A $};
\node() at (8,4.5) { \Huge $\bs 0$};
\node() at (2.5,2.) { \huge $\bs I_{\rho}$};

\draw [dotted] (0,0)--(0,-1);
\draw [dotted] (12,0)--(12,-2);
\draw [dotted] (6,0)--(6,-2);

\draw [dotted] (4,2)--(6,2);
\draw [dotted] (4,2)--(4,-2);

\draw [arr](0,-0.7) -- (12,-0.7);
\draw [arr](6,-1.5) -- (12,-1.5);
\draw [arr](6,-1.5) -- (4,-1.5);

\node() at (6.5,-1.) { \huge $\nu $};
\node() at (9,-2) { \huge $L =\nu-\rho$};
\node() at (5,-2) { \huge $\rho_A$};

\end{tikzpicture}
\caption{\label{shape} Structure of the parity-check matrix after diagonalization and reordering of columns}
\label{structure}
\end{figure}
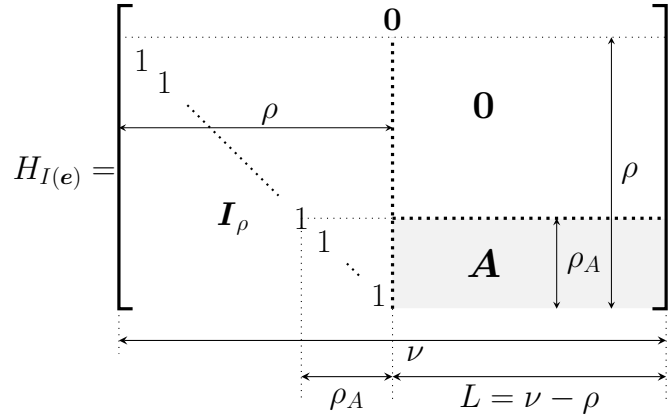

 If $L>0$, by using Guassian elimination
and  column and row permutations, the submatrix $\bldH_{I(\bs e)}$ can be represented in the form shown in
Fig.~\ref{structure}. Here, by $\rho_{A}$ we denote the rank of the system of linear equations left after Gaussian elimination. 
It is easy to see that the first  $\rho-\rho_A$ positions are uniquely determined and the other $L+\rho_A$ 
positions satisfy  
$\rho_A$ equations and cannot be determined uniquely. By assigning  arbitrarily values to $L$ of these  positions,
we uniquely determine  the remaining $\rho_A$ positions.    
A set  $I_{\rm AA}$   
of the corresponding  $L$ columns 
of the submatrix $A$   is  called  {\em arbitrarily assigned} (AA) positions.

\begin{definition}
{\it List erasure decoder} (LED) is a decoder for the BEC, which for a given input $\bs x$ with $\nu$ erased positions,
outputs a list ${\mathcal L}$ of codewords $\hat{\bs c}$ coinciding with the input word on all non-erased 
coordinates. 
%\footnote{\vit{I rewrote the definition of the list decoder. I was not sure why you defined the LED the way it was done. If I am missing something, please
%restore your definition.}}
\end{definition}

A possible implementation of  the LED is presented as Algorithm \ref{LED}. In this algorithm, a row is called \emph{a pivot} if it is chosen to eliminate nonzero elements in other rows in a position which we call \emph{a leader}.  

The algorithm combines   BP decoding over  the erasure channel and the Gaussian elimination steps. While there exists a row with one erasure, the BP decoding step is performed.  If there are no rows with one erasure, and there is a row which was not used as a pivot yet, the Gaussian elimination step with respect to the leader is performed.  Then, the algorithm switches back to  the BP decoding step. These alternating  steps are performed until neither rows with one erasure nor unused pivots are left.
\begin{algorithm}
\begin{algorithmic}[0]
\caption{\label{LED}LED algorithm for decoding of LDPC code on the BEC channel}
\Statex {\bf Input:} $\bs y = (y_1, \cdots, y_n) \in \{0,1,\phi\}^n$.
%\vspace{1mm}
\Statex {{\it Initialization}: \\
$I_{\rm AA} \gets \varnothing$; \\
$I(\bs e) \gets \{ i : y_i = \phi \}$ ; \\
$\nu \gets |I(\bs e)|$; \\
$\bs s \gets \bs y_{I^{\rm c}(\bs e)} \bldH^{\rm T}_{I^{\rm c}(\bs e)}$. 
%$\sigma$=number of checks with erasures.
}
%\vspace{1mm}
\Statex {\it Step 1:}  
\While {there is a check $j$ with one erased position $y_i$ }
\vspace{1mm}
{
\State
\parbox{7cm}  
{ 
$y_i \gets s_j$; \\
$I(\bs e) \gets I(\bs e) \backslash \{ i \}$; \\
$\nu \gets \nu-1$; \\%$\sigma \gets \sigma-1$, 
Update {\bs s};
}
\vspace{2mm}
\hspace{-5mm}
\If {$\nu=0$} {goto Step 3}; \EndIf 
}
\EndWhile
%\vspace{1mm}
\Statex {\it Step 2:} 
\If {there is a check $j$ with erasures not used as pivot}  \\ 
\vspace{-2mm}
\State 
\parbox{8cm}
{
\Statex{Select a check $j$ as a pivot; \\ Select $y_i = \phi$ as a leader;} %  \\
% \Comment {Gaussian elimination:}\;
\Statex{\it Gaussian elimination of row $j$:}\\
{Modify all checks which include position $i$; \\
Update $\bs s$;} 
\Statex {goto  Step 1.} 
\vspace{3mm}
}
%\vspace{-1mm}
\EndIf\\
\vspace{-3mm}
\Statex {\it Step 3}: $I_{\rm AA} \gets I(\bs e)$ , $\bs c \gets \bs y$.
%\vspace{1mm}
\Statex {{\it Step 4}: {\bf Return} $\bs c$ and $I_{\rm AA}$ .}  
\end{algorithmic}
\end{algorithm}
\section {Analysis of  the LED algorithm}
\label{sec_Analysis}
As it is mentioned in Section \ref{sec_Prelim}, we consider the Gallager ensemble of random $(J,K)$-regular LDPC codes.  
A random  $ r \times n$ parity-check matrix  which determines an $(J,K)$-regular LDPC code from this ensemble can be represented in the following form: 
 \begin{IEEEeqnarray}{C}
\bldH=\left(\begin{array}{c}
\bldH_{1}\\
\bldH_{2}\\
\vdots\\
\bldH_{J}
\end{array}
\right)\IEEEeqnarraynumspace
\label{polynomial2}
\end{IEEEeqnarray}
where  
\begin{IEEEeqnarray}{C}
\bldH_1=\left(\begin{array}{cccc}
{\bs 1}^{K}&{\bs 0}^{K}&\dots&{\bs 0}^{K}\\
{\bs 0}^{K}&{ \bs 1}^{K}&\dots&{\bs 0}^{K}\\
\vdots&\vdots&\ddots&\vdots\\
{\bs 0}^{K}& {\bs 0}^{K}&\dots&{\bs 1}^{K}
\end{array}
\right)\IEEEeqnarraynumspace
\label{polynomial3}
\end{IEEEeqnarray}
is of size $M\times n$, $M=r/J$, ${\bs a}^{K}=\underbrace{(a,a,...,a)}_{K}$  
and all the matrices $\bldH_{i}$, $i=2,3,\ldots,J$, are random permutations of the  columns of $\bldH_{1}$.  

In this section, we estimate the average size of the list $\mathcal L$ of candidate solutions in the system (\ref{eq:system}). Denote by $T$ a random variable, which represents this number of solutions.  A  set  of solutions of  
\begin{equation}
\bs z\bldH_{I(\bs e)}^{\rm T}=\bs 0
\label{coset}
\end{equation}
 represents a coset of solutions of  (\ref{eq:system}).  Next we analyze (\ref{coset}) instead of (\ref{eq:system}) since all cosets  have the same  number of solutions.    
Denote  by $ \bldH_{I(\bs e),1}$ a submatrix of $\bldH_{I(\bs e)}$ of size $M\times \nu$ consisting of its first $M=r/J$ rows. Next, we state the following lemma.%%%%%%%%%%%%%%%%%%%%%%%%%%
\lemma 
\label{lemma1}
Consider the Gallager ensemble of $(J,K)$-regular binary LDPC codes of length $n$ and  redundancy $r=nJ/K\gg 1$  over  the BEC with erasure probability $\varepsilon>0$. Then,  the conditional probability that  a  row in $\bldH_{I(\bs e),1}$  has zero weight  given that there are  
$\nu \gg 1$ erasures and $N_{0}<M$ rows have zero weight  does not grow with $N_{0}$ and is upper-bounded by 
$\left(1-\frac{\nu}{n}\right)^{K}$. 
\proof 
 Let $w_{i}$ denote the weight of  the $i$-th  row of $ \bldH_{I(\bs e),1}$, where $i=1,2,...,M$. Then, the probability $p_0$ that the $i$-th row has zero weight given that $\nu>0$ erasures occurred can be estimated as  
\begin{equation}p_0 \triangleq {\rm Pr}(w_{i}=0|\nu)= \frac{\binom{n-K}{\nu}}{\binom{n}{\nu}}=\frac { \binom{n-\nu}{K}  }
                           { \binom{n}{K}}   
                            \le 
                            \left( \frac{n-\nu}{n}  \right)^K \mbox{, } i=1,2,...,M.
\label{zero_pr}
\end{equation}

The conditional probability that $N_{0}<M$ rows  $i_{1},...,i_{N_{0}}$ of $ \bldH_{I(\bs e),1}$ have weight zero is given by: 
\begin{equation}
{\rm Pr}( w_{i_{1}}=0,...,w_{i_{N_{0}}}=0|\nu)={\rm Pr}(w_{i_{1}}=0|\nu)\prod_{j=2}^{N_{0}}
{\rm Pr}(w_{i_{j}}=0|{\bs w}_{i_{1}}^{i_{j-1}}={\bs 0},\nu),
\end{equation}

where ${\bs w}_{i_{1}}^{i_{j-1}}=(w_{i_{1}},w_{i_{2}},...,w_{i_{j-1}})$.

It is easy to see that  
\begin{equation}
{\rm Pr}(w_{i_{j}}=0|{\bs w}_{i_{1}}^{i_{j-1}}={\bs 0},\nu)=\frac{\binom{n-jK}{\nu}}{\binom{n-(j-1)K}
{\nu}}\le \frac{\binom{n-(j-1)K}{\nu}}{\binom{n-(j-2)K}
{\nu}} \, ,
\label{cond_prob}
\end{equation}
where the last transition is due to
\[\frac{\binom{n-jK}{\nu}}{\binom{n-(j-1)K}{\nu}} \frac{\binom{n-(j-1)K}{\nu}}{\binom{n-(j-2)K}
{\nu}}\le\left(\frac{n-jK}{n-(j-2)K}\right)^{\nu}=\left(1-\frac{2K}{n-(j-2)K}\right)^{\nu}<1.\]

We conclude that   the probability in  (\ref{cond_prob}) does not grow when the number of conditions increases, that is 
 the following chain of inequalities holds
\[
{\rm Pr}(w_{i_{j}}=0|{\bs w}_{i_{1}}^{i_{j-1}}={\bs 0},\nu)\le{\rm Pr}(w_{i_{j}}=0|{\bs w}_{i_{1}}^{i_{j-2}}={\bs 0},\nu)\le...\le {\rm Pr}(w_{i_{j}}=0| w_{i_{1}}= 0,\nu).\]
\begin{equation}
\le {\rm Pr}(w_{i_{j}}=0,\nu)=p_{0}. 
\label{ineq}
\end{equation}

\qed
 
In what follows, we present the main theoretical result of this work.
\theorem {Consider the Gallager ensemble of $(J,K)$-regular binary LDPC codes of length $n$ and  redundancy $r=nJ/K\gg 1$ over the BEC with the erasure  probability $\varepsilon>0$}. If there are   $\nu \gg1$ erasures,  then the ensemble average list size in LED, ${\rm E}[T|\nu]$,  is upper-bounded by
\[{\rm E}\left[T|\nu\right]\le 2^{\nu-r}\left( 1+\left(1-\frac{\nu}{n}\right)^{K}\right)^{r}.\]
\proof
{ 
Consider a random vector \[\bs s_{1}={\bs z}H_{I(\bs e),1}^{\rm T}.\] 
 Let $ I_{j}=\{i_{1}, i_{2}, \dots,i_{j}\}$, $j\le M$,  be an arbitrary set of  $j$ indices. Denote by $\bs s_{i_{1}}^{i_j}$   a subvector of $j$ components of the vector $\bs s$  indexed by $I_{j}$.  
 
If $\nu$ erasures occurred, then the probability that the random vector $\bs z$ is a solution of  
the system  $\bs z\bldH_{I(\bs e),1}^{\rm T}=\bs 0$
can be represented in the following form
\begin{equation}
{\rm Pr}\left(  {\bs z}H_{I(\bs e),1}^{\rm T}=\bs 0|\nu  \right)={\rm Pr}(\bs s_{1}=\bs 0|\nu)={\rm Pr}(s_{i_{1}}=0|\nu)\prod_{j=2}^{M}
{\rm Pr}(s_{i_{j}}=0|{\bs s}_{i_{1}}^{i_{j-1}}={\bs 0},\nu),
\label{main}
\end{equation}
where ${\bs s}_{i_1}^{i_{j-1}}=(s_{i_1},s_{i_2},\dots,s_{i_{j-1}})$.

For the choice of  a random vector $\bs z$ and a random parity-check  matrix from the Gallager ensemble,
the probability of a zero syndrome  component $s_{i}$ is
\begin{equation}{\rm Pr}(s_{i}=0|w_{i},\nu)=\left \{\begin{array}{ll}
1, & w_{i}=0\\
1/2, & w_{i}>0\\
\end{array}
\right.  
\label{scomp}, 
\end{equation}
where $i\in \{1,2,\dots,M\}$.
It follows from (\ref{scomp}) that
\[{\rm Pr}(s_{i}=0|\nu)={\rm Pr}(s_{i}=0|w_{i}=0,\nu){\rm Pr}(w_{i}=0|\nu)\]
\begin{equation}+{\rm Pr}(s_{i}=0|\nu,w_{i}>0)(1-{\rm Pr}(w_{i}=0|\nu)).
\label{synd_eq}
\end{equation}
By substituting (\ref{scomp})  into (\ref{synd_eq}), and by applying Lemma \ref{lemma1},  we obtain
\begin{equation}{\rm Pr}(s_{i}=0|\nu)=\frac{1+{\rm Pr}(w_{i}=0|\nu)}{2}\le \frac{1+\left(1-\frac{\nu}{n}\right)^{K}}{2}.
\label{onerow}
\end{equation}

In what follows, we show that 
\[{\rm Pr}(s_{i_{j}}=0|{\bs s}_{i_{1}}^{i_{j-1}}={\bs 0},\nu)\le {\rm Pr}(s_{i_{j}}=0|\nu).\]

Consider the probability ${\rm Pr}(s_{i}=0|s_{j}=0,\nu)$, $i,j\in \{1,2,\dots M\}$, $i\ne j$. By using the arguments similar to those in 
(\ref{onerow}), it is easy to obtain
\begin{equation}{\rm Pr}(s_{i}=0|s_{j}=0,\nu)=\frac{1+{\rm Pr}(w_{i}=0|s_{j}=0,\nu)}{2}.
\label{synd_cond}
\end{equation}

The conditional probability in the RHS of  (\ref{synd_cond}) can be represented as 
\[
{\rm Pr}(w_{i}=0|s_{j}=0,\nu)=\sum_{w_{j}=0}^{K}{\rm Pr}(w_{i}=0|s_{j}=0,w_{j},\nu){\rm Pr}(w_{j}|s_{j}=0,\nu)
\]
\begin{equation}
=\sum_{w_{j}=0}^{K}{\rm Pr}(w_{i}=0|w_{j},\nu)\frac{{\rm Pr}(s_{j}=0|w_{j},\nu){\rm Pr}(w_{j}|\nu)}{{\rm Pr}(s_{j}=0|\nu)}. 
\label{rhs}
\end{equation}
Substitution of  (\ref{scomp}) into (\ref{rhs}) yields
\begin{equation}
{\rm Pr}(w_{i}=0|s_{j}=0,\nu)=\frac{{\rm Pr}(w_{i}=0|w_{j}=0,\nu){\rm Pr}(w_{j}=0|\nu)+\sum_{w_{j}=0}^{K}{\rm Pr}(w_{i}=0|w_{j},\nu){\rm Pr}(w_{j}|\nu)}{2{\rm Pr}(s_{j}=0|\nu)}.
\label{sw_eq}
\end{equation}
In (\ref{sw_eq}), we took into account that ${\rm Pr}(w_{i}=0|s_{j}=0,w_{j},\nu)={\rm Pr}(w_{i}=0|w_{j},\nu)$.

Since  ${\rm Pr}(w_{i}=0|\nu)$ does not depend on $i$, from (\ref{sw_eq}), by using Lemma \ref{lemma1} and  (\ref{onerow}), we obtain  
\[
{\rm Pr}(w_{i}=0|s_{j}=0,\nu)={\rm Pr}(w_{i}=0|\nu)\frac{{\rm Pr}(w_{i}=0|w_{j}=0,\nu)+1}{2{\rm Pr}(s_{j}=0|\nu)} 
\]
\[\le {\rm Pr}(w_{i}=0|\nu)\frac{{\rm Pr}(w_{i}=0|\nu)+1}{2{\rm Pr}(s_{j}=0|\nu)}={\rm Pr}(w_{i}=0|\nu).
\]
From the last inequality and (\ref{synd_cond}), we conclude that 
\[{\rm Pr}(s_{i}=0|s_{j}=0,\nu)\le {\rm Pr}(s_{i}=0|\nu).\]
By using  similar  arguments it is easy to show that  
\[{\rm Pr}(s_{i_j}=0|{\bs s}_{i_1}^{i_{j-1}},\nu) \le {\rm Pr}(s_{i_j}=0|\nu)\le \frac{1+\left(1-\frac{\nu}{n}\right)^{K}}{2}.\]
Then, from (\ref{main}) we obtain
\[p_{1}\triangleq {\rm Pr}\left(  {\bs z}H_{I(\bs e),1}^{\rm T}={\bs 0}\right)\le  {\rm Pr}(s_{i_j}=0|\nu)^{M}\le \frac{\left(1+\left(1-\frac{\nu}{n}\right)^{K}\right)^{M}}{2^{M}}.
\]
%\begin{equation}
%\end{equation}

%\begin{equation} p_{1}\le \frac{N_{1}}{2^{\nu}}=\left(\frac{1+p_{0}}{2}\right)^{M}. \label{strip_prob}\end{equation}

Next, consider the submatrices  $ \bldH_{I(\bs e),i}$, $i=2,3,...,J$, consisting of rows 
$(i-1)M+1,..., iM$, respectively. Recall that the strips are obtained by the independent random permutations.
If $\nu$ erasures occurred then the probability that a random vector $\bs z$ is a solution of the system $\bs z \bldH_{I(\bs e)}^{\rm T}=\bs 0$  is upper-bounded as
\[{\rm Pr}(\bs z\bldH_{I(\bs e)}^{\rm T}=\bs 0|\nu)\le p_{1}^{J}= \frac{\left(1+\left(1-\frac{\nu}{n}\right)^K\right)^r}{2^r}.\]
%where the last transition is due to  (\ref{zero_pr}).

Notice that  analogous  to the derivations in \cite{gallager} we ignore the fact that parity-checks  of  $ \bldH_{I(\bs e)}$  are linearly dependent. If $\nu$ is large enough (i.e. grows linearly with $n$) then a number of linearly
dependent rows in $ \bldH_{I(\bs e)}$  is of order  $J$ and can be neglected.

Given that $\nu$ erasures occurred, we introduce  a random variable $\chi(\bs z)$ which is equal to 1 if $\bs z$ is a solution of $\bs z \bldH_{I(\bs e)}^{\rm T}=\bs 0$, and is equal to 0 otherwise. More formally,  
\[\chi(\bs z)=\left \{\begin{array}{ll}
1, & \bs z \bldH_{I(\bs e)}^{\rm T}=\bs 0\\
0, & \bs z \bldH_{I(\bs e)}^{\rm T}\ne \bs 0\\
\end{array}
\right. \; . 
\]

Then, the average list size ${\bf  E}[T|\nu]$ (given that there are $\nu$ erasures)  is equal to the average number  of vectors $\bs z$ which are solutions of  $\bs z \bldH_{I(\bs e)}^{\rm T}=\bs 0$, namely 
\begin{equation}
{\bf  E}[T|\nu]=\sum_{\bs z}{\bf E}[\chi(\bs z)|\nu]=\sum_{\bs z}{\rm Pr}(\bs z\bldH_{I(\bs e)}^{\rm T}=\bs 0|\nu)\le2^{\nu}\frac{\left(1+\left(1-\frac{\nu}{n}\right)^K\right)^r}{2^r}
 \label{sol} \; .
\end{equation} 
\qed
}
\corollary 
{Denote by $\alpha=\nu/r$ the normalized number of erasures. The  asymptotic exponent 
of the  list size is determined by  
\begin{eqnarray}
\varphi(\alpha,J,K)& =&
\lim_{r \to \infty}\frac{{\bf E}[L|\nu]}{r}\le \lim_{r\to \infty} \frac{{ \log_2\bf E} [T|\nu] }{r}\nonumber \\ 
&\le&  \alpha -1+\log_2 \left(1+\left(1-\alpha\frac{J}{K}\right)^K        \right) \; . 
\end{eqnarray}}
It is interesting to find a critical (largest) value of $\alpha$, such that $\varphi(\alpha,J,K)=0$ or, in other words, to find the relative number of erasures such that the average list size does not exceed 1.

%For random equiprobable\footnote{\vit{Does "equiprobable" means that the entries in the parity-check matrix are chosen Bernoulli($1/2$)?}}
% codes $\alpha=1$.
\begin{table}
\centering
\caption{\label{T1} Example of critical values of $\alpha=\nu/r$ }
\begin{tabular}{|c|c|c|c|c|c|c|}
 \hline
Rate    & 1/5   & 1/4     &  1/2   &  5/8 &  \multicolumn{2}{c|} {3/4}\\ \hline
$(J,K)$   & (4,5)   &(3,4)  &(4,8)& (3,8)& (3,12)&(4,16)\\ \hline
$\alpha$ &0.9995&0.994 &0.994&0.975&0.944&0.984 \\ \hline
\end{tabular}
\end{table}

%\end{document}

We expect that for sparse matrices it holds  $\alpha<1$. Examples of critical values of $\alpha$ for 
some code rates $R=1-J/K$ and for some values of $K$ are shown  in Table \ref{T1}.
We can see that $\alpha$ is close to~1 even for rather sparse parity-check matrices.
This suggests  that the allowable fraction of erasures $\nu/n$ can be chosen close to $1-R$. 

\vspace{-3mm}

%\section{Conversion of the AWGN channel into a BEC }
\section{Conversion of decoding over an AWGN channel into decoding over a BEC}
\label{Selection}
In \cite{fang2010bp}, the main criteria for bits to be erased is  the number of unsatisfied parity checks and low bit reliability values. The authors present therein  a set of thresholds which depend both on the code structure and on the channel signal-to-noise ratio (SNR). The corresponding bit is erased if the number of unsatisfied checks and the reliability value exceed the chosen thresholds.    

We use a different  strategy to transform the original problem of decoding over an AWGN channel  into a decoding problem over a BEC. By taking  into account that only $g/4$ iterations of  BP decoding can be considered independent, we analyze bit reliability values  obtained after $g$ iterations.  This allows us to avoid overestimating the reliability values. An overview of various techniques for processing BP reliability values for their further use in the near-ML decoding  can be found in \cite{baldi2014hybrid}. In our approach, first we calculate the minimum absolute values of bit reliability values (over $g$ iterations). Next, we sort them in the increasing  order. The $L_{1}$ least reliable bits are erased.      

 After erasing the $L_{1}$ least reliable  bits, we introduce  $L_2$ additional erasures by using a set of masks ${\mathcal M}= \{\mathcal M_i\}_{i=1,...,N}$, where
$\mathcal M_i\subset\{1,2,...,n\}$, and $|\mathcal M_i|=L_2$ for $i=1,2,\dots,N$. We use pseudo-random pre-selected binary sequences of length $2L_2$ and weight $L_2$ as masks. In our simulations we used codewords of the first order Reed-Muller code $[2^{m},m,2^{m-1}]$,  $m=\log_{2}L_{2}+1$, as such masks.
The masks are applied to the next  $2L_2$ least  reliable entries (after $L_1$ positions have already been erased).
This step in the algorithm is similar to the bit flipping step in the decoding algorithms such as \cite{valembois2004box,fossorier1995soft}. 
The choice of the parameters $L_{1}$ and $L_2$  depends on the code length and the code rate. 
In our simulations, we have chosen the total number of erasures $L_{1}+L_{2}$ to be:
\[L_{1}+L_{2}=\alpha(1-R)n \, , \]
where $\alpha \in [0.94, 1.07]$.   
We found empirically that for the rate $R \in \{1/3, 1/2, 2/3\}$, for any code length, we can choose $5 \le N \le 10$ and $L_2 \in [0.15 n, 0.18n]$.  

The BP-LED decoding can also be applied to decoding of nonbinary  LDPC codes over extensions of the Galois fields. It was found experimentally that for the $(2,4)$-regular LDPC code of length 16  over $GF(2^{8})$ (128 bits) constructed in \cite{Dolecek}, the choice  $\alpha=1.4$ gives the best FER performance. In our experiments, $L_{2}$ was chosen to be $0.15n$.
 
\section{LED-based  algorithm for an AWGN channel \label{ledbased}}

In this section, we show how the LED can be used for decoding of LDPC codes on an AWGN  channel  with binary phase shift keying  (BPSK) signaling. Let {${\mathcal C} = \{\bldc_j\}_{j = 0,1, \cdots, 2^k-1}$  be a binary $[n,k,d_{\min}]$ LDPC code. Assume that $\mathcal C$  is used with BPSK and coherent detection to communicate over an AWGN channel. The binary code symbol 
$c_{i}\in \{0,1\}$ is mapped onto the signal $v_i=(2c_i-1)\sqrt{E_{\rm s}}$, $i=1,2,\dots,n$, where $E_{\rm s}$ is the signal energy.  In the sequel we assume that  $E_{\rm s}=1$. Thus the codewords ${\bs c_{j}}=(c_{1}^{(j)},c_{2}^{(j)},\dots,c_{n}^{(j)})$, $j=0,1,\dots,2^{k}-1$, are mapped onto bipolar sequences  $\bs v_{j}=(v_1^{(j)},v_2^{(j)},...,v_n^{(j)})$. Assume that ${\bs v}_{0}$ is transmitted.
Then the discrete-time received signal is $ {\bs r} = {\bs v}_{0} + {\bs n} $, where the noise vector $\bs n$ consists of independent zero-mean Gaussian random variables  with variance $\sigma^{2}=N_{0}/2$. The SNR per information bit is denoted by $E_{\rm b}/N_{0}$ = $E_{\rm s}/(N_{0}R)$, where $ R = k/n$ is the code rate.

Assume that  $\bs v=(v_1,v_2,...,v_n)$ and  $\bs r=(r_1,r_2,...,r_n)$ are the  transmitted and the received vectors, respectively. 
Let $J_{\max}\le 2^{L}$ be  the maximal number of  allowed candidate solutions, and 
$\mu(\cdot)$ be a decoding metric. 
We use the  Euclidean distance between the channel output $\bs r$ and
$2 \bs c - 1$ as the decoding metric $\mu(\bs c)$, where $\bs c$ is a candidate codeword. 
Alternatively, we can maximize the scalar product of $\bs r$ and $2\bs c-1$.

The new BP-LED decoding algorithm is presented as Algorithm \ref{LEDalg}. 

In Algorithm \ref{LEDalg}, first,  $\nu=L_1+L_2$ input symbols are erased (see below). 
 Then  LED is used for correcting of  ${\rho}$ erasures and for making  a list $I_{\rm AA}$ of the AA positions. 
\smallskip

The high-level idea of the proposed decoding algorithm is as follows. The algorithm  consists of the three main steps:
\begin{enumerate}
\item{BP decoding. }
\item{In case of BP decoding failure, the unreliable positions are erased and LED is applied.}
\item{In case of  the LED failure to find a unique solution, exhaustive search over a list of $J_{\max}$ 
candidate codewords is carried out.}
\end{enumerate}    
\medskip

These three steps are  implemented as the following three subroutines in Algorithm \ref{LEDalg}. 
\begin{itemize}
\item  
$(\hat{ \bs v},\bs x)= \mbox{\sc bpdecod}(\bs r)$,  where $\hat{ \bs v}$ and $\bs x$ are 
vectors of hard decisions and of symbol reliabilities, respectively, produced by the BP decoder. 
As it is mentioned above,  in order to avoid overestimates due to cycles in the Tanner graph, $\bs x$ is computed as 
the minimum of the absolute values of the symbol reliabilities in the first $g$ iterations.

\item 
$(\bs c, I_{\rm AA})= \mbox{\sc LED}(\bs \xi)$, where $\bs \xi$ is a vector $\bs v$ with 
zeros on $\nu$ 
erased  positions, and $\bs c$ is a vector of hard 
decisions with erasures in  $L$ AA positions.
%\footnote{\vit{This sentence is not well formulated. Not clear what the list $\mathcal L$ contains. Would it be wrong to have the list $\mathcal L$ containing all possible codewords (solutions) that were found?}}.
Function LED is as discussed  in Section~\ref{BEC}.

\item
$\bs c_{j}= \mbox{\sc candidate} (\bs c,I_{\rm AA},j)$. This subroutine generates the $j$-th candidate codeword
from the full list of solutions of~(\ref{eq:system}). This is done by constructing a list $\mathcal W$ of 
$J_{\max}$ binary words of length $L=|I_{\rm AA}|=\log_{2}|\mathcal L|$  ordered according to the ascending order of their
weights.
%\footnote{\vit{Not clear: is the length of the list $L_{\max}$ or $L$? What is the difference between $L_{\max}$ and $L$?}}. 
Then, the $j$-th candidate is obtained by flipping AA bits in positions determined by the ones of 
the $j$-th element in $\mathcal W$. 
\end{itemize}
\begin{algorithm}
\begin{algorithmic}[0]
\caption{\label{LEDalg}Algorithm for  decoding of LDPC code on the AWGN channel}
\Statex {\bf Input:}  the vector of LLRs $\bs r = (r_1, \dots, r_n) \in \mathbb R^n$.  
\Statex Let $\mu_{\rm opt} \gets \infty$. \\
\vspace{1mm}
{\it Step 1:} \\
$(\hat{\bs v}, \bs x)=\mbox{\sc bpdecod}(\bs r)$;  
${\hat {\bs c}}=({\hat{\bs v}}+1)/2$;
\If {$\hat{\bs c}\bldH^{\rm T}=\bs 0$ }  goto Step 6; \EndIf \\
\vspace{1mm}
{\it Step 2:} 
\Statex
$\bs \xi \gets \hat{ \bs v}$ with zeros on 
$L_1$ least reliable positions in $\bs x$.
\For{ $i=1$ to $N$}
{
%\begin{description}
\vspace{0.2cm}
\State	\parbox{8cm} {{\it Step 3:}  \\ Use mask ${\mathcal M}_i$ to erase $L_2$  
	non-erased positions in $\bs \xi$. }%Replace nonerased positions of $\bs x$ by hard decisions. }
%
%Denote $I_i$ the set of the erased positions, 
%	$\bs x$ corresponding hard decisions sequence with erasures, $x_i={\rm sign}(y_i), i\notin I_i$, 
%	$x_i=0, i\in I_i$}

\vspace{0.2cm}
\State \parbox{8cm} {{\it Step 4:} \\
	$({\bs c}, I_{\rm AA}) \gets \mbox{\sc LED}(\bs \xi)$;\\
	Initialize AA positions in $\bs c_{0}$ by hard decisions from $\bs r$.
%based on 
%	$\bs y$: $c_i=1$, if $y_i>0$, $i\in \mathcal L$ 
}  
\vspace{1mm}
\State \parbox{8cm} {\it Step 5:}
%	\item[Step 5]
	 \For{ $j=1$ to $J_{\max}$}
		{ 
\vspace{0.1cm}
\State	\parbox{7cm}{Compute codeword $\bs c'$ from $\bs c$ and $\bs c_{j-1}$;} 
		\If {$\mu(\bs c') < \mu_{\rm opt}$ } let $\hat{\bs c} \gets \bs c'$; $\mu_{\rm opt} \gets \mu(\bs c')$; \EndIf
%\vspace{0.3cm}

	%\State	\parbox{7cm}{
\hspace{0mm} Generate next candidate  
$
\bs c_{j}=\mbox{\sc candidate}  ( \bs c, I_{\rm AA}, j )  
$;
%}

%\State	\parbox{7cm} {Choose next combination of inversions from the ordered set of allowed combinations  }
%		%of length $L=|\mathcal L|$
%\vspace{0.3cm}
%\State	\parbox{7cm} {Apply inversions to $\bs c$ and reconstruct estimated codeword $\bs c'$}
%\vspace{0.3cm}
%\State	Compute metric $\mu=\mu(\bs c')$ 
%\vspace{0.3cm}
%		\If {$\mu<\mu_{\rm opt}$ } let $\hat{\bs c}=\bs c', \mu_{\rm opt}=\mu$ \EndIf
		}
	\EndFor
%\end{description}
}
\EndFor \\
{\it Step 6:} Return  $\hat{\bs c}$.

\end{algorithmic}
\end{algorithm}
%\end{document}
% \vspace{-2mm}
\section{ Discussion and Simulation results}
\label{sec_Discussion}
In this section, we compare experimentally  the FER performance of  BP decoding (with 50 decoding iterations) with that of  the BP-LED decoding. We also compare the experimental results with the tightened theoretical upper bounds  for binary random linear codes and for the Gallager ensembles of   binary reqular  LDPC codes  and binary images of nonbinary regular LDPC codes  under ML decoding. 
Specifically, we simulate the rate $R=1/2, 1/3, \mbox{ and }2/3$ binary irregular LDPC codes of length $n=576$ and the rate $R=1/2$ binary $(4,8)$-regular  LDPC code of length $n=96$.       
We also simulate the LED-based  decoding for the rate $R=1/2$ nonbinary LDPC code of length $n=16$ over $GF(2^{8})$ \cite{Dolecek} and compare its  performance with the FER performance of the generalized BP decoding \cite{davey1998low} of the same code, and with the corresponding theoretical upper and lower bounds. 

The experimental results are shown in Fig. \ref{fernb}. While simulating the BP- LED  decoding for nonbinary codes over  extensions  of the binary field we recomputed probabilities of bit values via probabilities of symbol values of $\GF(q)$, $q=2^{m}$, as follows

\begin{eqnarray*}
{\rm Pr} (b_{i,j}=0) & = & \sum_{ \lambda\in{\rm GF}(q) \; : \; b_{i,j}=0} {\rm Pr} (b_i=\lambda), \quad \mbox{$j=1,2,...,m$} \, , \\
{\rm Pr} (b_{i,j}=1) & = & 1-{\rm Pr} (b_{i,j}=0) \; ,
\end{eqnarray*}
where ${\rm Pr} (b_i=\lambda)$ is the probability of the $q$-ary symbol $b_i$ to be equal $\lambda$,
${\rm Pr} (b_{i,j}=l), \; l \in \{0,1\}$, $ \; j=1,2\ldots,m$, is the probability that the $j$-th bit in the binary representation of $b_i$ is equal to $l$, and \\ ${b_{i}=(b_{i,1},...,b_{i,m})}$ is the binary representation of $b_i$.

All parity-check matrices of the simulated  binary irregular LDPC codes  were  constructed by using the optimization technique  in \cite{Boch2016}. The parity-check matrix of the $(4,8)$-binary regular code was obtained by reducing modulo 6 the degree matrix of the double Hamming LDPC code in \cite{bocharova2011double}.
In order to show that the chosen codes are on a par  with the  best LDPC codes used in communication standards, 
the FER performance of   BP decoding for the standard WiMAX codes of rates $R=1/2$ and $R=2/3$ is presented in Figs. \ref{fer12} and \ref{fer23}. To facilitate the  low complexity encoding, the degree matrices of all simulated codes of length $n=576$ have the so-called bi-diagonal form. The FER performance of the $n=96$ LDPC code was compared with the  FER performance of near-ML BEAST-decoding \cite{beast}. 

In the  simulations of the  binary LDPC codes, the parameters  $\alpha$ and $\beta$ were  optimized over the range $[0.94,1.07]$ and  $[0.15,0.18]$, respectively.  Among the $\alpha(1-R)n$ erased positions, $(\alpha(1-R)-\beta)n$  positions were selected according to the reliabilities, estimated  by the BP decoding, and $\beta n$ positions were selected pseudo-randomly from the next $2\beta$ less reliable positions. Two values of the sizes of the pseudo-random sets,  $N=5$ and $N=10$,  as well as two values of the list sizes,  $2^{8}$ and $2^{16}$, were simulated. All the simulations were run until at least 100 LED-decoding block errors occurred. It turns  out that for all codes, except for the rate $R=2/3$ irregular LDPC code, two sets of parameters: $N=5$ and the list size $2^{16}$; $N=10$ and the list size $ 2^{8}$,--- provide  approximately the same coding gain of the LED-based algorithm with respect to the BP decoding. For the rate $R=2/3$ code, the LED-based algorithm with a larger list size yields a slightly lower FER. In nonbinary case, the parameter $\alpha=1.4$ is selected, the number of trials is $N=10$ and the list size is $2^{8}$.
  
In order to compare the FER performance of the BP-LED decoding of both regular and irregular LDPC codes  with the FER performance of  ML decoding, in Figs. \ref{fer12}--\ref{ferhamm} we present an upper bound  (\ref{TSB}) computed for both the random linear codes and for the $(J,K)$-regular  LDPC codes from the Gallager ensemble.  In case of irregular LDPC codes, the upper bound is computed for the parameters $J$ and $K$  chosen to be equal to the average number of ones in columns and rows of the parity-check matrix, respectively. The Shannon lower bound (\ref{Shannon}) is presented in the same figures as well.
% \end{document}  
From the presented results, we conclude  that the coding gain is higher for the regular code than for the optimized irregular LDPC codes. We also observe that  for binary codes the coding gain  grows with $E_{\rm b}/N_{0}$.  
For higher rates, the FER performance of the LED-based algorithm is closer to the  corresponding upper bound on  ML decoding performance than for lower rates. It is easy to see that the coding gain compared to the FER performance of   BP decoding for the rate $R=1/2$ WiMAX code is significant. However,  the rate $R=2/3$  WiMAX code  has the same FER performance as 
the optimized LDPC code. For the $(4,8)$-regular  LDPC code of length $n=96$, the coding gain of  the new algorithm with respect to the BP decoding is approximately two times smaller than the coding gain obtained by the near-ML BEAST decoding.  
In nonbinary case, the coding  gain does not grow with $E_{\rm b}/N_{0}$.  Such behavior of the FER performance can be explained by using LED in combination with generalized BP decoding which is superior to conventional BP decoding. Higher efficiency of  generalized BP decoding reduces the gap in the FER performance of ML and BP decoding. As a result near-ML decoding provides a smaller coding gain than that for the binary LDPC codes.     

The analysis of computational complexity of Steps~4 and~5 of  Algorithm \ref{LEDalg} is presented in \cite{bocharova2016low}. Although for a general linear code, the computational complexity of Step 4  would be a cubic function of the code length $n$, the empirically  observed average decoding time for LDPC codes grows near-linearly with the code length.  
The computational complexity of Step 5 is proportional to the list dimension $L$, that is, it grows linearly  with the  code length as well. 

Although complexity grows near-linearly, the algorithm loses efficiency for 
large $n$ since a  typical number $L$ of AA positions grows as well. To maintain the decoding 
efficiency, parameters $N$ and $J_{\max}$ should also be increased, which 
leads to impractically high computational time  for lengths above 2000.   

\begin{figure}
\begin{center}
\includegraphics[width=120mm]{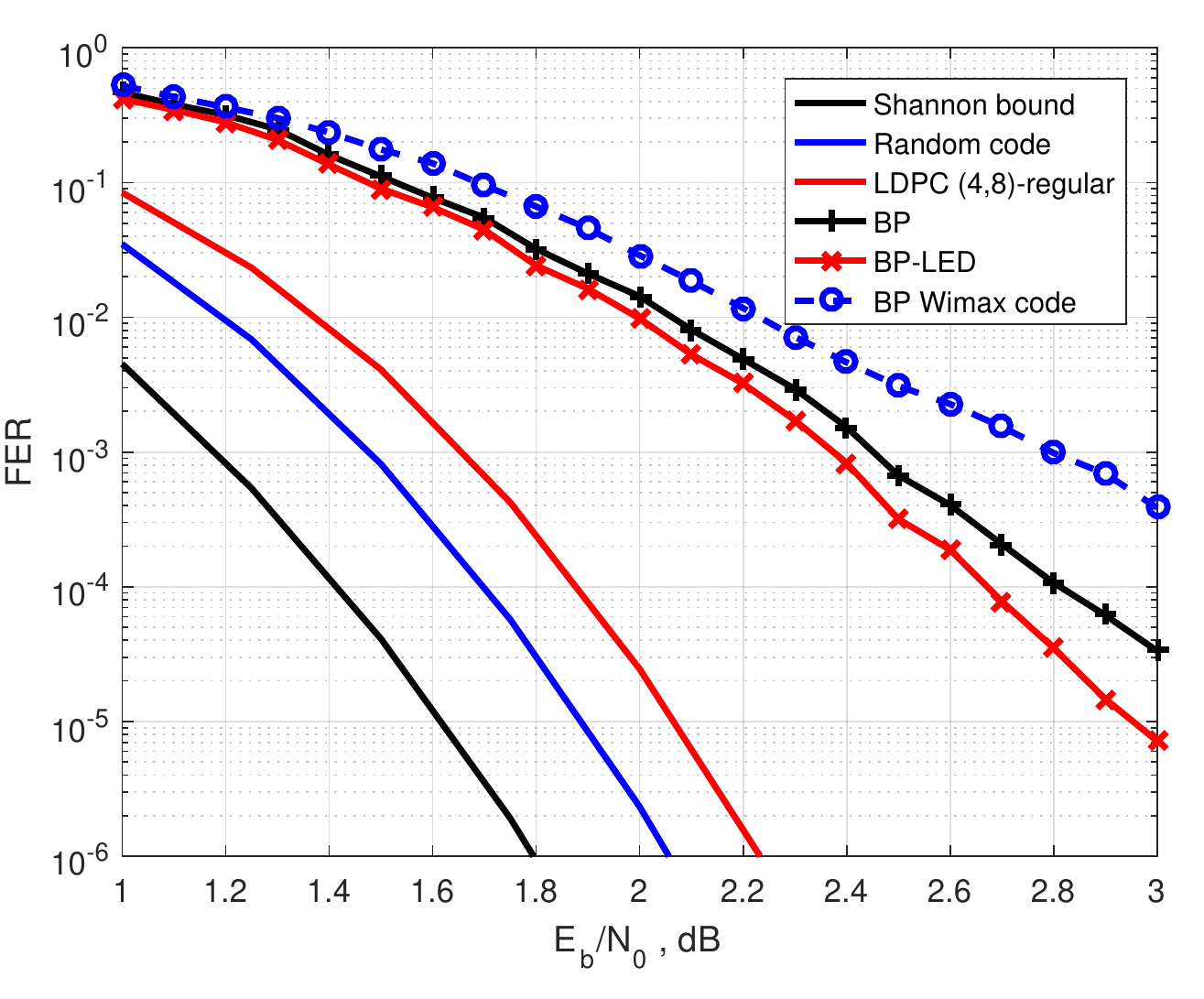}   
\caption{Bounds and FER performance for BP and near-ML decoding of $R=1/2$ 
 irregular  LDPC codes of length $n=576$, 
 where the following notations are used: ``Shannon bound'' 
 denotes  the bound (\ref{Shannon}), ``Random code'' denotes the 
 bound (\ref{TSB})   computed for ensemble of linear codes, 
 ``LDPC (4,8)-regular'' denotes the bound  (\ref{TSB})  computed 
 for the Gallager ensemble of  the (4,8)-regular  LDPC codes.   
\label{fer12} }
\end{center}
\end{figure}
\begin{figure}
\begin{center}
\includegraphics[width=120mm]{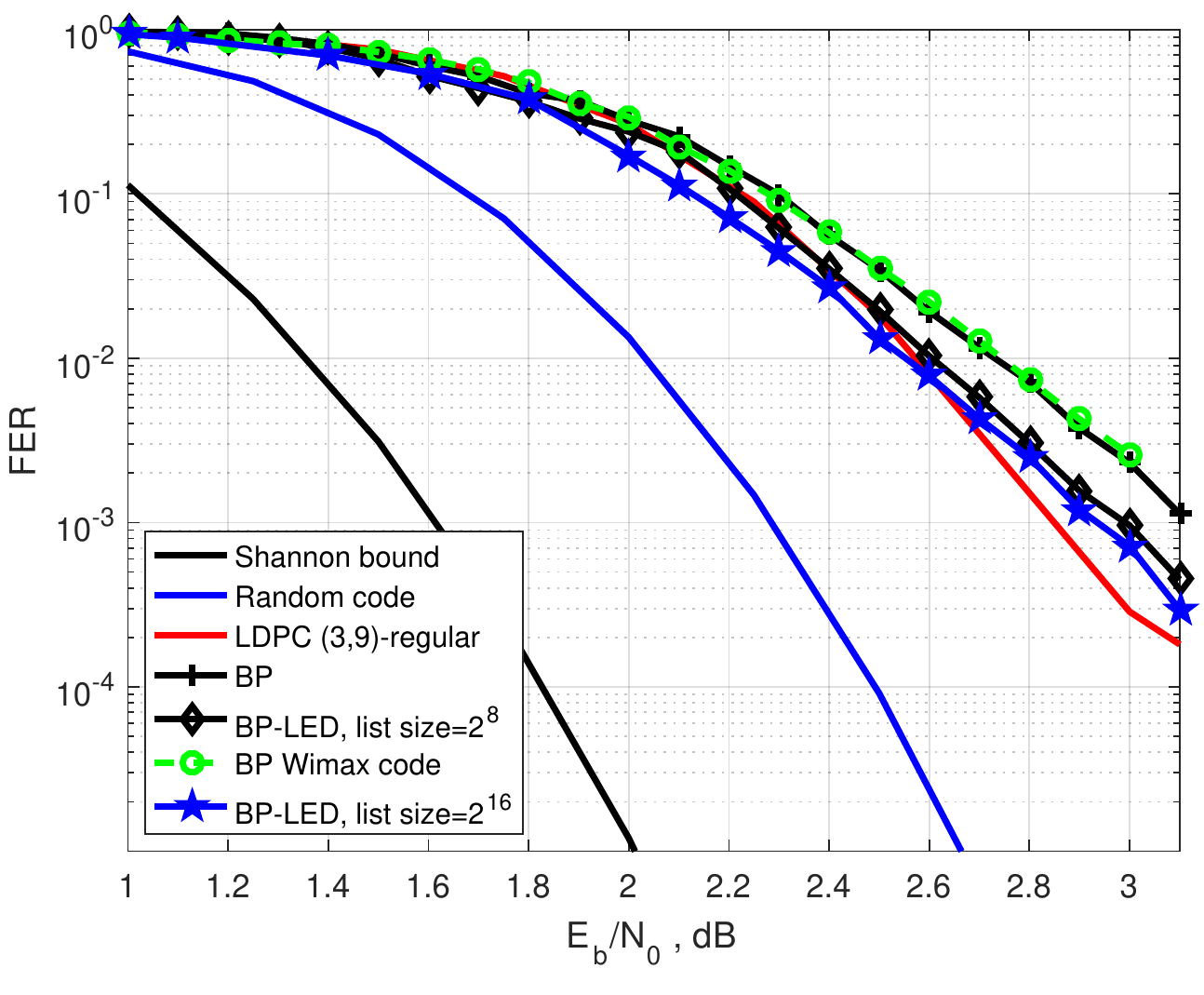}   
\caption{Bounds and FER performance for BP and near-ML decoding of $R=2/3$ irregular  LDPC codes of length $n=576$, where the following notations are used: ``Shannon bound'' denotes  
the bound (\ref{Shannon}), ``Random code'' denotes the 
bound (\ref{TSB})   computed for ensemble of linear codes, ``LDPC (3,9)-regular'' 
denotes the bound  (\ref{TSB})  computed for the Gallager ensemble of  the (3,9)-regular LDPC codes.   
 \label{fer23} }
\end{center}
\end{figure}
\begin{figure}
\begin{center}
\includegraphics[width=120mm]{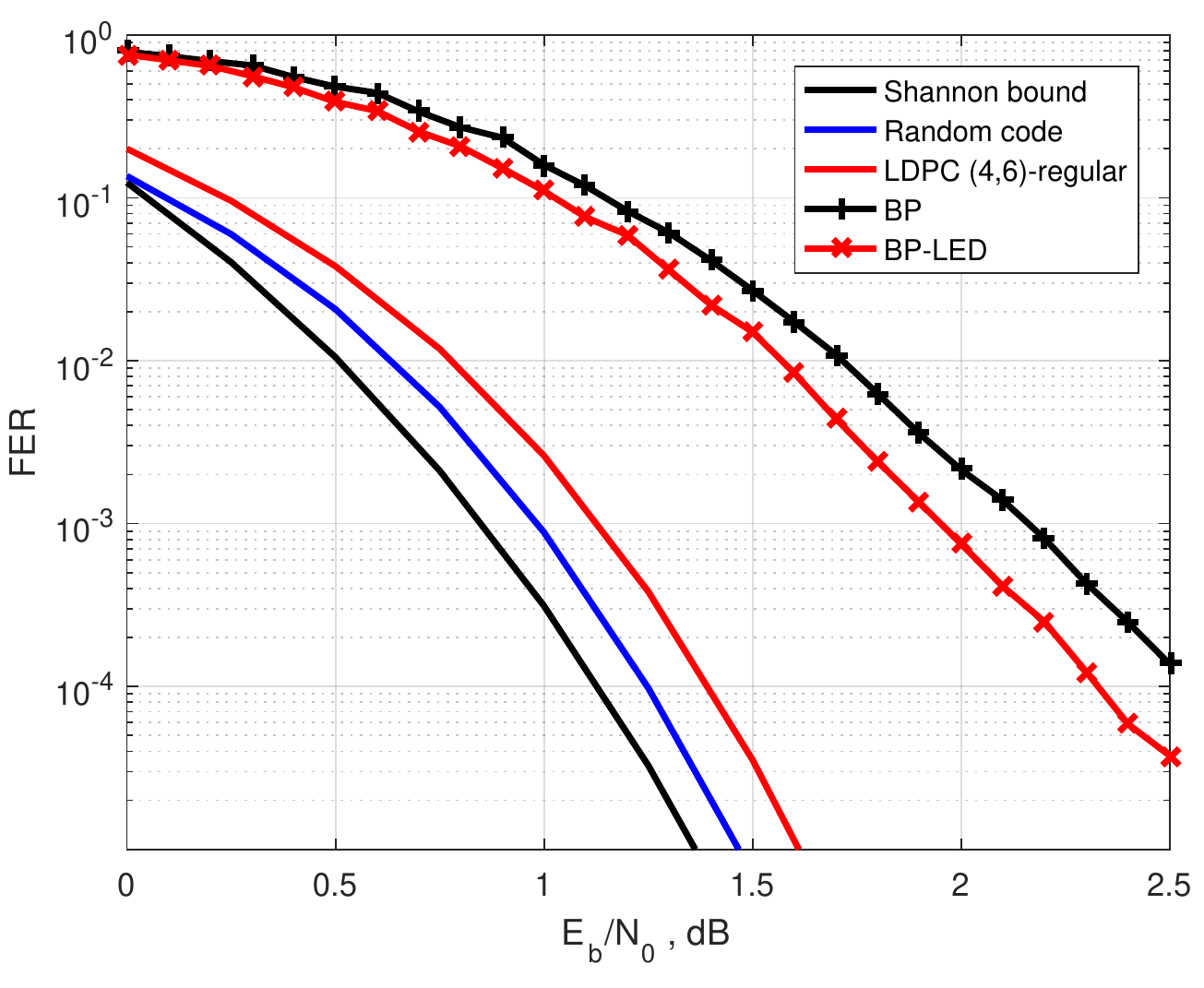}   
%\caption{Bounds and FER performance for BP and near-ML decoding of $R=1/3$ \bor{irregular} LDPC codes of length $n=576$ }
%\label{fer13}
\caption{Bounds and FER performance for BP and near-ML decoding of $R=1/3$ irregular  LDPC codes of length $n=576$, where the following notations are used: ``Shannon bound'' denotes  
the bound (\ref{Shannon}), ``Random code'' denotes the 
bound (\ref{TSB})   computed for ensemble of linear codes, ``LDPC (4,6)-regular'' 
denotes the bound  (\ref{TSB})  computed for the Gallager ensemble of the (4,6)-regular LDPC codes.   
 \label{fer13} }

\end{center}
\end{figure}
\begin{figure}
\begin{center}
\includegraphics[width=120mm]{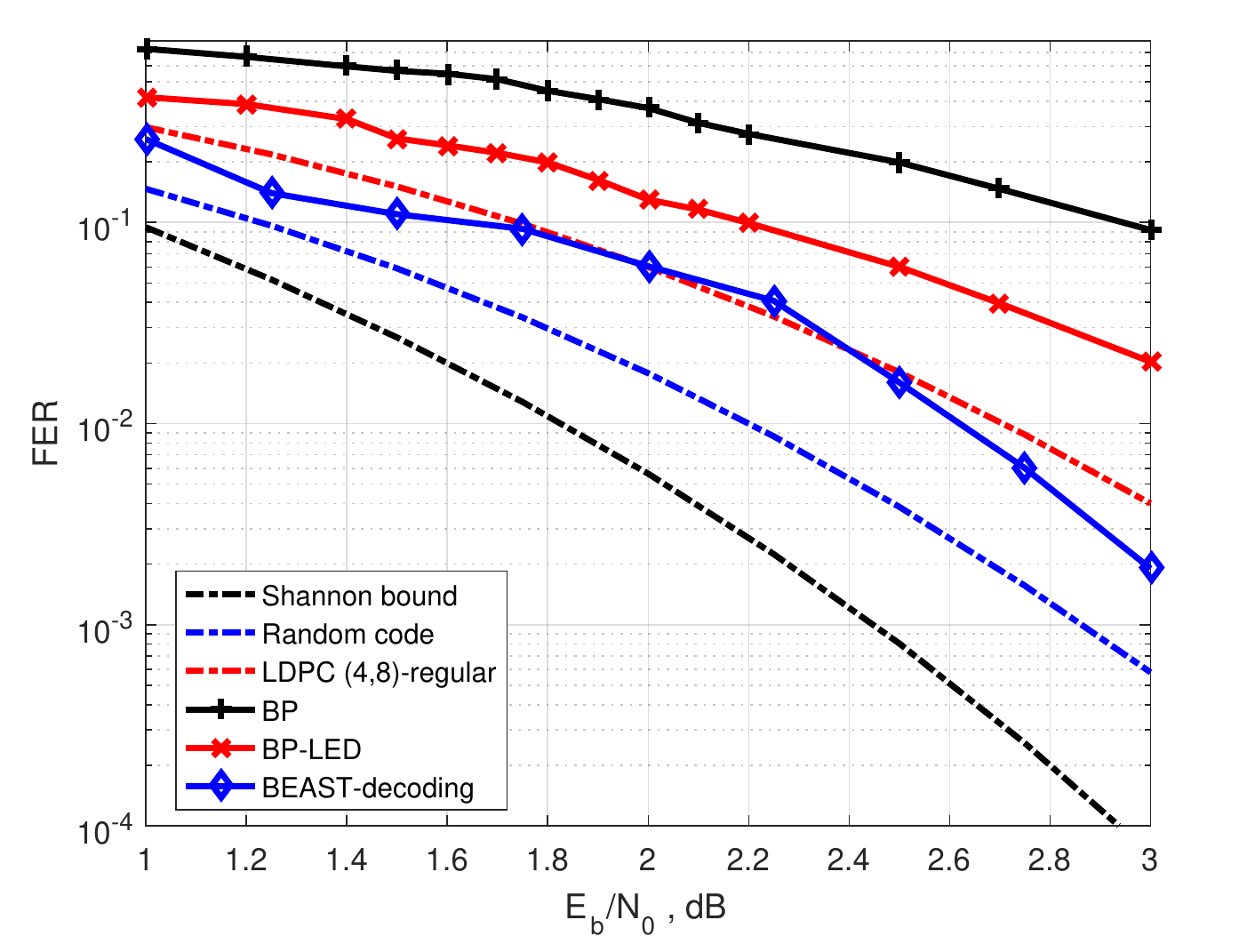}   
\caption{Bounds and FER performance for BP and near-ML decoding of $R=1/2$ $(4,8)$-regular  LDPC codes of length $n=96$, where the following notations are used: ``Shannon bound'' denotes  
the bound (\ref{Shannon}), ``Random code'' denotes the 
bound (\ref{TSB})   computed for ensemble of linear codes, ``LDPC (4,8)-regular'' 
denotes the bound  (\ref{TSB})  computed for the Gallager ensemble of the (4,8)-regular LDPC codes. 
\label{ferhamm}}
\end{center}
\end{figure}

%\end{document}

\begin{figure}
\begin{center}
\includegraphics[width=120mm]{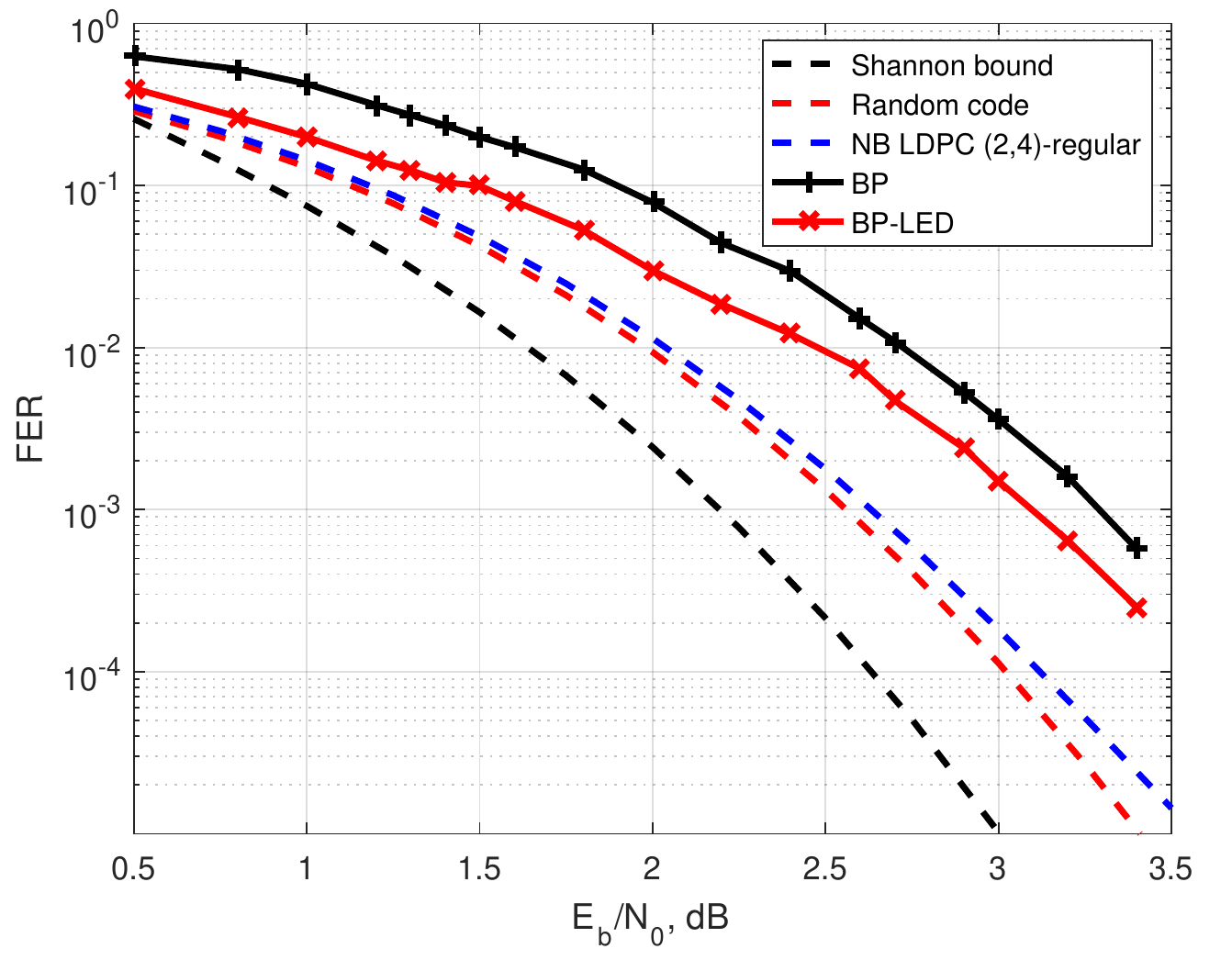}   
\caption{Bounds and FER performance for BP and near-ML decoding of $R=1/2$ nonbinary  LDPC code of length $n=16$ over $GF(2^{8})$ \cite{Dolecek}, where the following notations are used: ``Shannon bound'' denotes  
the bound (\ref{Shannon}), ``Random code'' denotes the 
bound (\ref{TSB})   computed for ensemble of linear codes, `` NB LDPC (2,4)-regular'' 
denotes the bound  (\ref{TSB})  computed for the Gallager ensemble of  binary images of the (2,4)-regular NB LDPC codes.
\label{fernb}}
\end{center}
\end{figure}

\section{Conclusion}
A new algorithm for near-ML decoding of LDPC codes  over the AWGN channel is proposed and analyzed.   
The new algorithm as well as  BP decoding are simulated for both the regular and irregular binary QC LDPC codes of several rates and  for the binary image of a short nonbinary LDPC code over the extension of the binary Galois field. The FER performance for binary and nonbinary LDPC codes is compared to the improved union-type upper bound on the error probability  based on precise  coefficients of the average spectra for the Gallager ensembles of binary regular LDPC codes and binary images of nonbinary regular LDPC codes, respectively. The coding gain of the new decoding algorithm strongly  depends on  the communication scenario. Decoding performance close to the performance of  ML decoding  is demonstrated for the  high rate LDPC codes as well as for the short length LDPC codes.

\bibliographystyle{IEEEtran}
\bibliography{listdec}
\end{document}